\documentclass[12pt]{iopart}
\usepackage{graphicx}
\usepackage{graphics}
\usepackage{float}
\usepackage{color}

\usepackage{dcolumn}   
\usepackage{bm}        
\usepackage{amssymb}   

\begin{document}

\title[]{
Luneburg lens  waveguide networks
}
\author{M. M. Mattheakis$^{1,2}$, G. P. Tsironis$^{1,2}$, V. I. Kovanis$^{3}$
}

\address{
$^{1}$Department of Physics, University of Crete, P. O. Box 2208, 71003 Heraklion, Greece }
\address{
$^{2}$Institute of Electronic Structure and Laser, Foundation for Research and Technology-Hellas, N. Plastira 100,
Vassilika Vouton, GR-70013, Heraklion, Crete, Greece}
\address{
$^{3}$Air Force Research Laboratory, Sensors Directorate, Wright Patterson Air Force Base, Ohio 45433, USA}
\ead{\mailto{mariosmat@physics.uoc.gr}}
\ead{\mailto{gts@physics.uoc.gr}}
\ead{\mailto{vassilios.kovanis@gmail.com}}


 \begin{abstract}
We investigate  certain configurations of Luneburg lenses
that form light propagating and guiding networks. We study  single Luneburg
lens dynamics  and  apply the single lens ray tracing solution 
  to various arrangements of multiple lenses.  The wave propagating features of the Luneburg lens networks
  are also verified through 
  direct numerical solutions of Maxwell's equations.  We find that Luneburg lenses may form
efficient waveguides for light propagation and guiding.  The additional presence of nonlinearity 
improves the focusing characteristics of the networks.
\end{abstract}

\maketitle

\section{Introduction}
Gradient Index (GRIN) metamaterials are formed through spatial variation of the index of refraction and lead to
enhanced light manipulation in a variety of circumstances. These metamaterials provide natural
means for constructing various types of waveguides and other optical configurations that guide and focus light in
specific desired paths.  Different configurations have been tested experimentally while   
the typical theoretical approach uses Transformation Optics (TO) methods to cast the original inhomogeneous index problem to an equivalent one in a deformed space \cite{TO_meta, leonh_TO_GL, leonh_1, pendry, chen, plasmons2}.  While this approach is mathematically elegant, it occasionally hides the 
intuition obtained through more direct means.  Furthermore, a general, continuous GRIN waveguide may be hard to analyze in  more
elemental units and relate its global features to these units.  In the present work, we adopt precisely this latter avenue, viz. 
attempt to construct waveguide structures that are seen as lattices, or networks, of units with specific features.  This is a 
"metamaterials approach", where specific properties of the "atomistic" units are inherited as well as expanded in 
the network.

The "atomic" unit of the network is a Luneburg lens (LL); the later is a spherical construction where the index or refraction
varies from one, in its outer boundary, to $\sqrt{2}$ in the center with a specific functional dependence on the lens radius \cite{luneburg, general_solution}.
Its basic property, in the geometrical optics limit, is to focus parallel rays on the spherical surface on the opposite side of the
lens \cite{plasmons2, luneburg,  general_solution, falco, analytical_solution, barba_2, barba_3, cloak, plasmons1}. This feature makes LL's quite 
interesting for applications since the focal surface is predefined for parallel rays of any 
initial angle.  While the rays traverse the lens, they suffer variable  deflections depending on their distance from
the optical axis ray, leading to a point image on the lens surface.  This property of the spherical LL is also shared by
its cylindrical equivalent formed by long dielectric cylinders while the light wave-vector impinges perpendicularly to
the cylindrical axis.  This geometry turns the problem into a two dimensional one, constructed for any plane that cuts the
LL cylinder perpendicular to its axis.  The work in this article focuses on exactly this type of cylindrical LLs and,
as a result, our approach is strictly two dimensional \cite{luneburg, general_solution, analytical_solution}.

In the present work we analyze the ray trajectories in a single LL and subsequently we use this analysis in 
waveguides build exclusively through LLs.  Specifically, in the next section we present analytical approaches for the  
trajectories  traversing a LL \cite{TO_meta, leonh_TO_GL, leonh_1, luneburg, general_solution, analytical_solution, lagrangian, stavroudis, modern_optics, born, parazzoli} and 
derive explicitly a  solution expressing the ordinate of the trajectory as a function of the abscissa in 
Cartesian coordinates. In section 3 we introduce the main point of the present work, viz. the formation of Luneburg Lens Waveguides (LLW)  through the arrangement of multiple LLs \cite{cloak} in geometrically linear or
bent configurations.  The LLWs are investigated through dynamical maps stemming from  the analytical solution for  the single LL.  In section 4 we investigate the same LLWs but using a direct solution of Maxwell's equations and find compatibility with the results of ray tracing approach  of the previous section. Additionally, we also use nonlinearity in the index of
refraction and find better focusing characteristics \cite{barba_3, barba_4}.  Finally, in section 5 we conclude.

\section{Single Luneburg lens}
\subsection{Ray tracing solution}
 
The general problem we have at hand is light propagation in an inhomogeneous isotropic medium
with an index of refraction $n(\vec r)$, where $\vec r$ is the position vector.  Although this
problem has been tackled though various approaches, for the purposes of our problem we need 
the exact solution for propagation through a single LL, which is itself an inhomogeneous medium.
What we need is a ray tracing solution for a single LL that, in turn, will be used for providing
the basic element in a map approach for the ray dynamics in LL networks.

Given the recent interest in LLs we present below three ways for
obtaining  the ray tracing solution for the single LL; these explicit solutions from the
different approaches may be useful in
other LL studies as well. The first two methods (section 2.2 and Appendix A.1 respectively) are based on the Fermat principle  for optical path or travel time minimization~\cite{TO_meta,luneburg,lagrangian,stavroudis,modern_optics,born} while the last one (Appendix A.2)  uses a geometrical optics approach to Helmholtz wave equation~\cite{TO_meta,luneburg,modern_optics,born, parazzoli,helmholtz}.  In all cases we focus on the specific, cylindrically symmetric, LL index of refraction $n(\vec r)$~ \cite{luneburg,general_solution,analytical_solution,parazzoli}

\begin{equation}
\label{lune_index}
n(r)=\sqrt{2-(\frac{r}{R})^2}
\end{equation}
 where $R$ is the radius of the lens while $r$ is the radial distance from the center in the interior of the lens, i.e. $r \leqslant R$.  The LL is supposed to be embedded in a medium with index equal to one,
 leading to a continuous change in the index variation as a ray enters the lens.
 
 \subsection{Quasi 2D ray solution}
 
 The optical path length $\mathcal{S}$ of a ray from a point A to B is given by \cite{TO_meta,luneburg,lagrangian,stavroudis,modern_optics,born}

\begin{equation}
\label{2.1}
\mathcal{S}=\int_A^B n ds
\end{equation}

In polar coordinates the arc length  is 
$ds=\sqrt{1+r^2\dot\phi^2}dr$, with
 $\dot \phi \equiv d\phi / dr$ where the coordinate $r$ is considered a "generalized time". As
 a result Fermat's variational integral of Eq. (\ref{2.1}), becomes 
 \begin{equation}
 \label{2.3}
\mathcal{S}=\int_A^B n(r)\sqrt{1+r^2\dot{\phi}^2} dr
\end{equation}
and leads to the Optical Lagrangian \cite{TO_meta,lagrangian,stavroudis}

 \begin{equation}
 \label{2.4}
\mathcal{L}(\phi,\dot \phi, r)=n(r)\sqrt{1+r^2 \dot \phi^2} 
\end{equation}
The shortest optical path is obtained through minimization of the integral of Eq. (\ref{2.3}) and may be
found through the solution of the Euler-Lagrange equation for the Lagrangian of Eq. (\ref{2.4}), viz.
\begin{equation}
\label{2.5}
\frac{d}{dr} \frac{\partial \mathcal{L}}{\partial \dot{\phi}}=\frac{\partial \mathcal{L}}{\partial \phi}
\end{equation}
Since the Lagrangian (\ref{2.4}) is cyclic in $\phi$, we have
$\partial \mathcal{L} / \partial \phi=0$ and thus
$\partial \mathcal{L} / \partial \dot{\phi}=\mathcal{C}$
where $\mathcal{C}$ is a constant. We thus obtain for the 
 Lagrangian of Eq. (\ref{2.4}) \cite{TO_meta,lagrangian,stavroudis}

\begin{equation}
\label{differential}
\frac{n(r)r^2 }{\sqrt{1+r^2 \dot{\phi}^2}}\dot{\phi}=\mathcal{C}
\end{equation}
This is a nonlinear differential equation describing the trajectory $r(\phi )$ of a ray in the interior of 
LL. Replacing the term $\dot \phi\equiv d\phi / dr$ and solving for $d\phi$, we obtain a first integral of motion, i.e.

\begin{equation}
\label{2.7}
\int{d\phi} = \int{\frac{\mathcal{C}}{r \sqrt{n^2r^2-\mathcal{C}^2}}dr}
\end{equation}
 The expression of Eq. (\ref{2.7}) holds for arbitrary index functions $n(r)$.  Using the specific LL refractive
index function of Eq. (\ref{lune_index}) we evaluate the integral (\ref{2.7}) and obtain the ray tracing equation for a 
single LL as follows
 
\begin{equation}
\label{m1.9}
r(\phi) =  \frac{\mathcal{C'} R}{ \sqrt{1-\sqrt{1-\mathcal{C'}^2}~\sin(2(\phi+\beta))} }
\end{equation}
where $\mathcal{C'}$ and $\beta$ are constants. This analytical expression may be cast in a direct Cartesian
form for the $(x,y)$ coordinates of the ray; after some algebra we obtain

\begin{eqnarray}
\label{m1.10}
(1-T\sin(2\beta))x^2+(1+T\sin(2\beta))y^2 \nonumber \\
-2T\cos(2\beta) xy+(T^2-1)R^2=0
\end{eqnarray}
where $T$ and $\beta$ are constants. We note that Eq. (\ref{m1.10}) is an equation for an ellipse. This  result agrees  with the Luneburg theory and states that inside a LL light follows elliptic orbits.
 
The constants $T$ and $\beta$  of Eq. (\ref{m1.10}) are determined through the ray boundary (or "initial" conditions) and depend on the initial propagation angle $\theta$ of a ray that enters the lens at the point  $(x_0,y_0)$ lying on the circle at the lens radius $R$ \cite{lagrangian}. The entry point of the ray is at $(x,y)=R(\cos(\theta+\pi),\sin(\theta+\pi ))=-R(\cos\theta , \sin\theta )$. Substituting these expressions in Eq. (\ref{m1.10}) we obtain after some algebra
 
 \begin{equation}
\label{relation_A}
T=\sin(2\beta +2\theta)
\end{equation}

In order to determine both constants $T,\beta$ we need an additional relation connecting them. Taking the derivative of the Eq.(\ref{m1.10}) respect to  $x$ and regarding the fact that $dy/dx=\tan(\theta)$, where $\theta$ the initial propagation angle. In addition, using the labels $(x_0,y_0)$ for the initial ray point on the LL surface, we set $x=x_0$ and $y=y_0$ in Eq.(\ref{m1.10}) we solve for $T$ and obtain
 
 \begin{equation}
\label{relation_B}
 T = \frac{x_0+y_0 \tan(\theta)}{\tan(\theta)[x_0 \cos(2\beta)-y_0\sin(2\beta)]+[x_0 \sin(2\beta)+ y_0 \cos(2\beta)]} 
\end{equation} 

The Eqns. (\ref{relation_A}) and (\ref{relation_B}) consist of an algebraic nonlinear system expressing the
constants $T$ and $\beta$ as a function of the initial ray entry point in the LL at $(x_0 , y_0 )$ with angle $\theta$. 
Combining the Eqns. (\ref{relation_A}), (\ref{relation_B}) and after simplifications
we obtain 

\begin{equation} 
\label{sol_beta}
\beta=\frac{1}{2}(\tan^{-1}({x_0}/{y_0})-\theta)
\end{equation}
therefore, according Eq. (\ref{relation_A})
\begin{equation}
\label{sol_tau}
T=\sin(\tan^{-1}(x_0 /y_0 )+\theta)
\end{equation} 
 
Substituting now the Eqs.(\ref{sol_beta},\ref{sol_tau}) to Eq.(\ref{m1.10}) and solving for $y$, we obtain the equation
 
\begin{eqnarray}
\label{m1_11}
y(x)=\frac{ (2x_0 y_0+ R^2 \sin(2\theta))}{2x_0^2+(1+\cos(2\theta))R^2}x ~~+ \nonumber \\
\nonumber \\
+\frac{\sqrt{2}R y_0\cos(\theta)\sqrt{(1+\cos(2\theta))R^2+2x_0^2-2x^2} }{2x_0^2+(1+\cos(2\theta))R^2}- \nonumber \\
\nonumber \\
-\frac{x_0 \sin(\theta)\sqrt{(1+\cos(2\theta))R^2+2x_0^2-2x^2}}{2x_0^2+(1+\cos(2\theta))R^2} ~~~~~
\end{eqnarray}
where $x_0, y_0$ give the initial ray entry point and $\theta$ the initial propagation angle.

The Eq. (\ref{m1_11}) describes the complete solution of the ray trajectory through a LL; 
in the simplest case where all rays are parallel to the $x$ axis, the initial angle  $\theta=0$ and Eq. (\ref{m1_11}) 
simplifies to 
\begin{equation}
\label{m1_12}
y(x)=\frac{y_0}{x_0^2+R^2}(x_0 x+R\sqrt{R^2+x_0^2-x^2})
\end{equation}

We note that in order to determine the exit angle $\theta '$, i.e. the angle with  which each ray exits the lens,
we may take the arc tangent of the derivative of Eq. (\ref{m1_11}) wrt $x$ at the focal point on the surface of lens 
at $x=R\cos(\theta)$.  The solution of Eq. (\ref{m1_11}) may be used to study several configurations
of LLs; however special provisions are necessary in cases where the rays need to turn backwards.  
It is thus more practical to use parametric solutions where the ray coordinates $x,y$ are 
both dependent variables.  This approach is explained in the Appendix where the parametric solution
of Eq. (\ref{m3_solution}) is derived.


\section{Luneburg Lens Waveguides}

The analytical solutions for the single LL trajectory, viz.
Eqs.(\ref{m1_11}) or (\ref{m3_solution})  may be used in order to study analytically
the ray transfer through various configurations of LLs that form waveguides \cite{cloak}. Using the initial entry
point on the LL $(x_0 , y_0)$ as well as the initial ray angle $\theta$ we obtain through Eq. (\ref{m1_11})
the exit point $(x,y)$ and the associated exit angle $\theta '$.  We may thus form a mapping from $(x_0 , y_0 , \theta )$
to $(x , y, \theta ')$; further propagation in the surrounding medium is rectilinear while the entry to the next
LL is governed by a new initial entry point with angle equal to the previous exit angle.  The resulting ray may be traced easily through the map.

\begin{figure}
\centering
\includegraphics[scale=0.5]{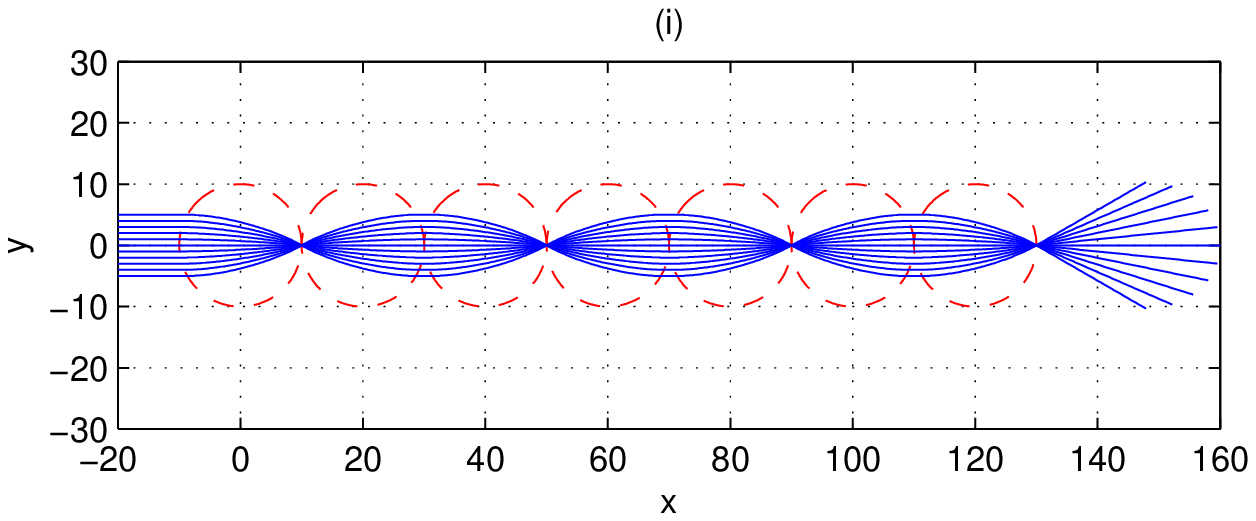}
\includegraphics[scale=0.5]{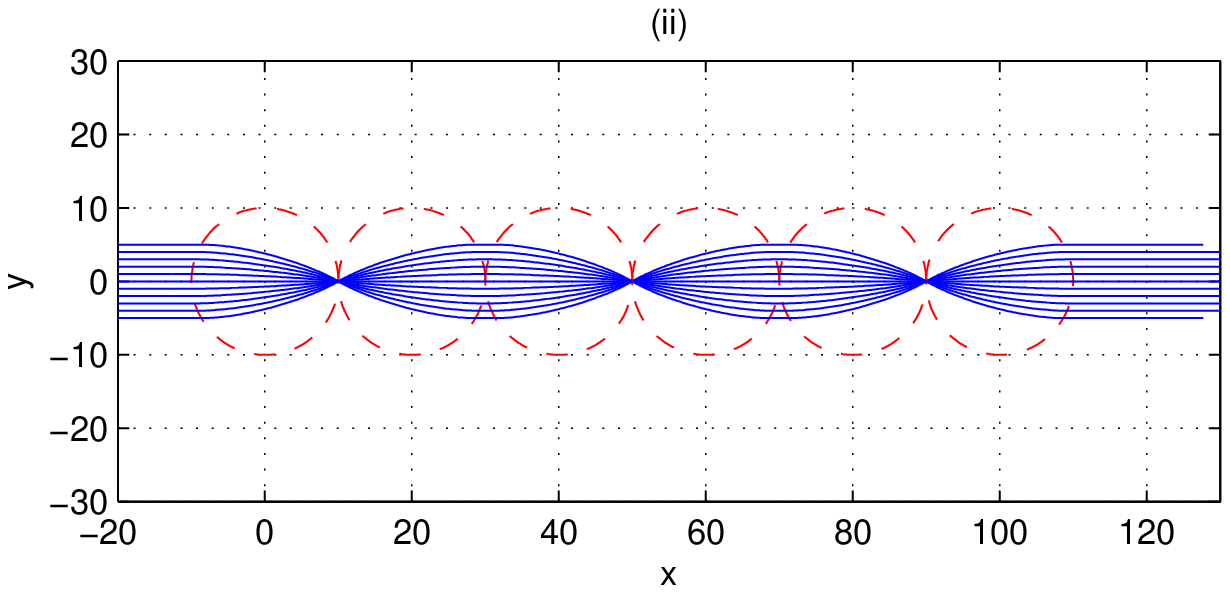}
\caption{The red dotted lines denote the arrangement of lenses. The blue lines show the ray tracing. Light is guided by Luneburg lenses across the linear network. (i) Arrangement with 7 lenses (left figure). We can see that all rays are focus on the last lens. (ii) Arrangement with 6 lenses (right figure). All rays exit in the
same mode as originally entered, i.e. parallel to the waveguide axis.}
\label{fig_ray_linear}
\end{figure}

A geometrically linear  arrangement of touching LLs on a straight line is shown in  in  Fig. \ref{fig_ray_linear}; this configuration forms a waveguide that channels the light through.  Depending
on the number of lenses even or odd number) the rays focus in the last LL surface or exit as they
entered respectively.  In both cases we sent a beam of rays parallel to the axis of symmetry of the
network.

\begin{figure}
\centering
\includegraphics[scale=0.6]{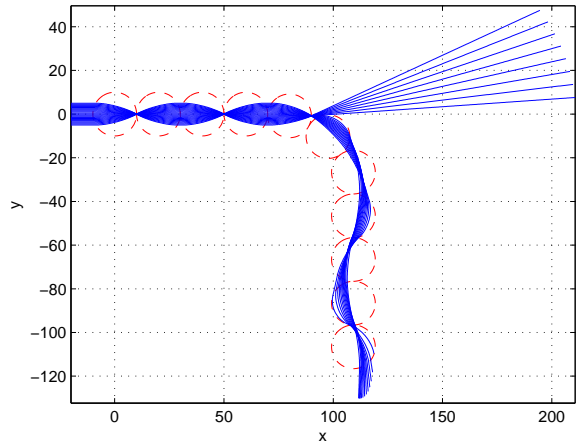}
\caption{A right angle waveguide is formed through a network of eleven Luneburg lenses. 
The ray-tracing is performed analytically using Eq. (\ref{m3_solution}). We observe efficient guiding
with a total bend of  $90^o$ degrees.}
\label{fig_ray_rectangular}
\end{figure}

In Fig. (\ref{fig_ray_rectangular}) we use the  ray tracing equations, Eqs.(\ref{m1_11}) or (\ref{m3_solution}) to study the propagation along curved arrangements
of LL's.  We observe the efficient channeling of the rays through the network that leads to a complete
ray banding at a right angle.  Some rays escape in the sharp bend, but generally the
guiding is very efficient for such a drastic change of direction.

\begin{figure}
\centering
\includegraphics[scale=0.6]{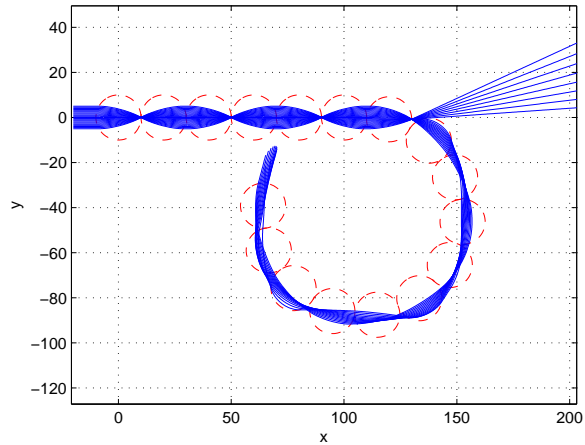}
\caption{A full circle bend waveguide formed through seventeen Luneburg lenses. The ray-tracing is
performed through the ray tracing methods. We see that Luneburg waveguides can guide  light
 in a complete circle bend trajectory; more complex paths are also possible.}
\label{fig_ray_circular}
\end{figure}

In   Fig. (\ref{fig_ray_circular}) we form a full circle bend waveguide through a sequence of LLs and follow the light propagation in the geometric optics limit.  We find  that light may propagate efficiently through a loop, signifying that arbitrary waveguide formation and guiding is possible.

The LL network cases presented (linear, right angle, curved) signify that LLs may be used as
efficient waveguides. Their advantage over usual dielectric guides is that light bending occurs
naturally through the LL properties while the outgoing light may be also focused, if so desired.
In bends there are naturally some losses that, in the geometric optics limit, may be estimated 
by comparing  the number of the incoming  to the outgoing rays, viz. $N_{IN}$) versus $N_{OUT}$
respectively. In the linear  arrangement of waveguides, as in Fig. \ref{fig_ray_linear},  the performance is perfect since $N_{IN}=N_{OUT}$. In the bend cases, such as in the right angle arrangement  of Fig. \ref{fig_ray_rectangular} as well as in the full circle bend waveguide of Fig. \ref{fig_ray_circular}, we find $N_{OUT}/N_{IN}=13/21=0.6$.  We note that the aforementioned losses depend also on the ray coverage of the initial lens as well as the sharpness of the bend; the losses can be reduced by manipulating appropriately these two factors. 

\section{Wave propagation in Luneburg lens networks}

\subsection{Propagation through LL waveguides}

In the present section we present results for the true wave propagation in networks of LLs.
The simulations we done both with a homemade FDTD based algorithm that was tested against COMSOL Multiphysics$^{\textregistered}$ software.  Both gave good agreement and, as a result, we show only the results obtained
from the latter.   In order to be more realistic,  we used Gaussian beams instead  of plane waves \cite{barba_3,barba_4}.  In the single LL case  the effect of the gaussian beam shape is
to shift the focus outside the LL surface, an effect that
may be compensated though nonlinearity ~\cite{barba_3,barba_4}. 
Specifically, we simulate all three arrangements shown previously, viz.  Figs. \ref{fig_ray_linear}, \ref{fig_ray_rectangular} and 
\ref{fig_ray_circular} and we calculate numerically the losses of the guiding. Subsequently, we introduce  nonlinearity and show that the Kerr effect improves guiding as well as focusing of the beam through Luneburg waveguide.

In order to setup our arrangements we use silica glass for the Luneburg lenses with refractive index variation  as in Eq.(\ref{lune_index}). In addition we use air for the outer medium with index of refraction $n=1$ \cite{barba_2,barba_3,plasmons1,plasmons2} with a monochromatic TM mode Gaussian EM wave. We perform all simulations in microwave regime with wavelength  $\lambda =1cm$. In all simulations the
radius of each LL  is ten times greater that the wavelength, i.e.  $R=10\lambda=10cm$. Finally, in the lattice edge we apply PML boundary conditions \cite{taflove}.

\begin{figure}
\centering
\includegraphics[scale=1]{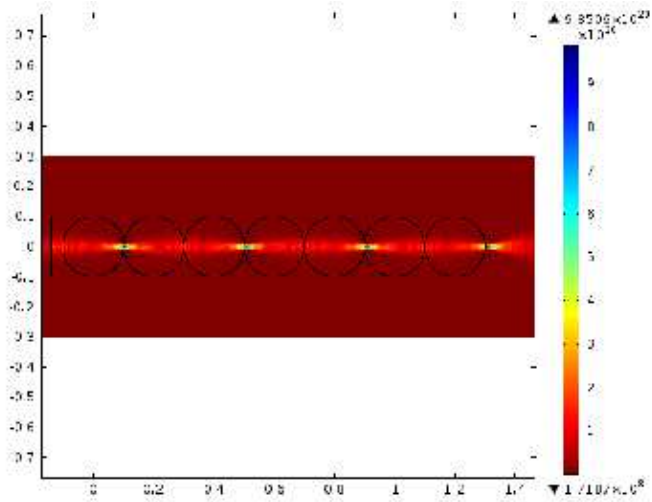}\\
\caption{ Linear Luneburg waveguide arrangement with seven lenses. Beam guiding and focusing on the surface of the last lens is observed. The result is in agreement with Fig. \ref{fig_ray_linear}(i). Furthermore, no losses are observed.}
\label{fig_comsol_linear_L7}
\end{figure}

\begin{figure}
\centering
\includegraphics[scale=1]{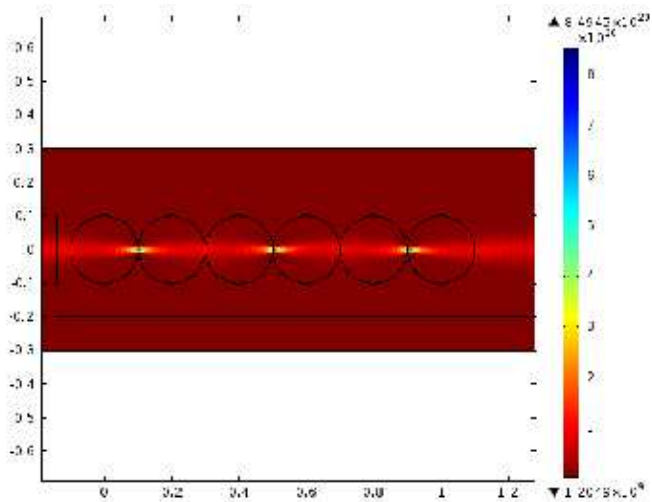}
\caption{Linear waveguide arrangement formed with six Luneburg lenses. The EM wave is guided through the linear network of lenses and the wave defocusses in the last lens. The result is in agreement with Fig. \ref{fig_ray_linear}(ii). Additionally, the guiding is almost perfect since the losses are practically zero.}
\label{fig_comsol_linear_L6}
\end{figure}

\begin{figure}
\centering
\includegraphics[scale=1]{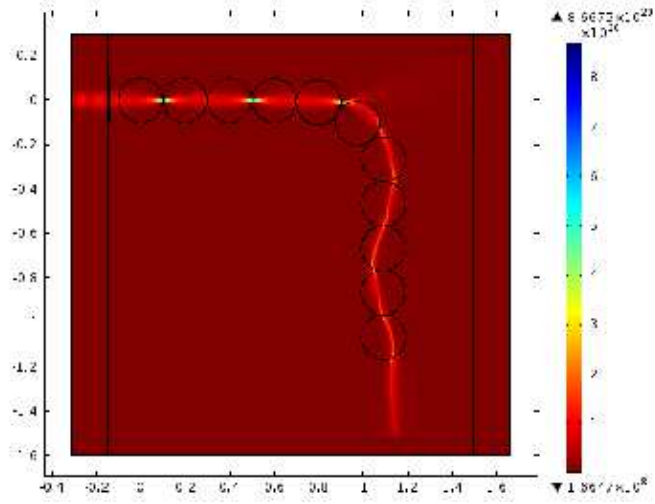}
\caption{A right angle bend waveguide is formed by eleven Luneburg lenses. The EM wave is guided and bends at right angle through a network of lenses. The losses in this case are approximately 27\% of the initial amplitude, i.e  $E_{OUT}/E_{IN}=0.73$.}
\label{fig_comsol_rectangular_L11}
\end{figure}

\begin{figure}
\centering
\includegraphics[scale=1]{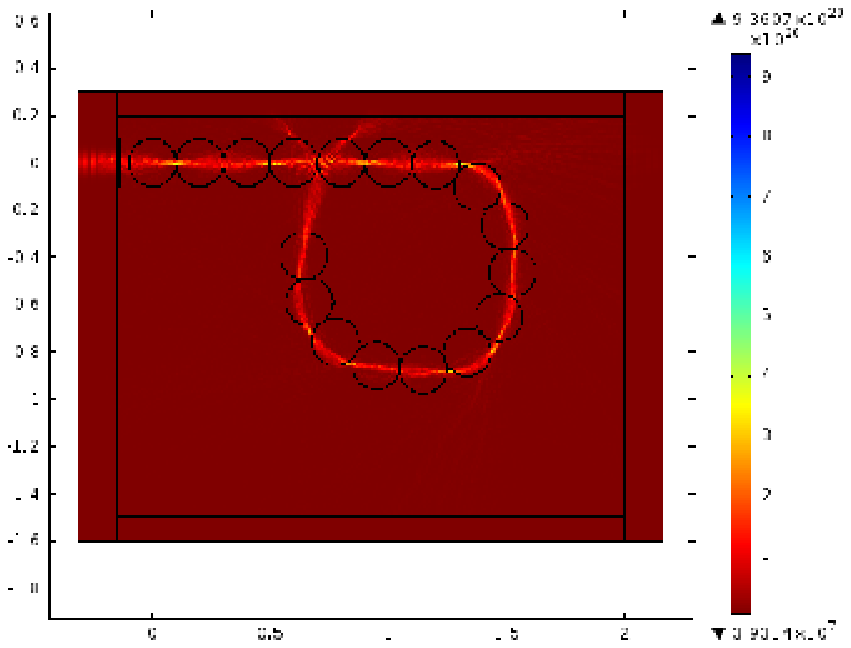}
\caption{ A full circle bend waveguide is formed by seventeen Luneburg lenses. The EM wave is guided through a circular orbit. The losses in this arrangement are approximately 30\% since  $E_{OUT}/E_{IN}=0.70$.}
\label{fig_comsol_circular_L17}
\end{figure}

In Figs. (\ref{fig_comsol_linear_L7}-\ref{fig_comsol_circular_L17}) we show the results for the full propagation of EM waves through LLs waveguides. We plot the steady state of the absolute value of the electric field. 

For the linear arrangement of LL waveguides we study a sequence with seven and six lenses, Figs.  \ref{fig_comsol_linear_L7} and \ref{fig_comsol_linear_L6} respectively. In the case with even number of lenses, as in Fig. \ref{fig_comsol_linear_L7}, the ray beam is guided and defocusses while exciting the
waveguide. On the other hand, in the case with odd number of lenses, as in Fig. \ref{fig_comsol_linear_L6}, the beam focuses in the last lens. The results obtained are
compatible with those obtained through the 
ray tracing methods of  Fig. \ref{fig_ray_linear}
The right angle bend is shown in  Fig. \ref{fig_comsol_rectangular_L11}, while in Fig.   \ref{fig_comsol_circular_L17} we have a full circle arrangement of LLs; these arrangements are the same as those analyzed in Fig. \ref{fig_ray_rectangular} and Fig. \ref{fig_ray_circular} respectively.   In all
cases studied, the numerical solution of Maxwell's equations is compatible with the findings obtained
through the  ray tracing map.

\subsection{Propagation losses}

In order to find how efficient is the guiding by LL waveguides we may calculate losses due to  propagation through LL networks. Subsequently, we calculate numerically the absolute value of the electric field in the first and in the last lens in each arrangement. Specifically, we calculate the
integrated field intensity at a cross section that is perpendicular to the propagate direction and is located
in the center of each lens and compare outgoing to incoming waves.
We find that in the linear case of Fig. \ref{fig_comsol_linear_L7},\ref{fig_comsol_linear_L6} that $E_{OUT}/E_{IN}=0.98$. As a result,  linear guiding is almost perfect. For the right angle bend waveguide of the Fig. \ref{fig_comsol_rectangular_L11} we find that $E_{OUT}/E_{IN}=0.73$ while  in the circular arrangement of the Fig. \ref{fig_ray_circular} $E_{OUT}/E_{IN}=0.70$. We  conclude that
guiding is quite efficient for such sharp bends. We note that 
losses can be reduced further through  using LLs with different radii as well
as more   efficient geometries that have less sharp curvatures.

\subsection{Nonlinear Kerr effect in LL waveguides}

\begin{figure}
\centering
\includegraphics[scale=1]{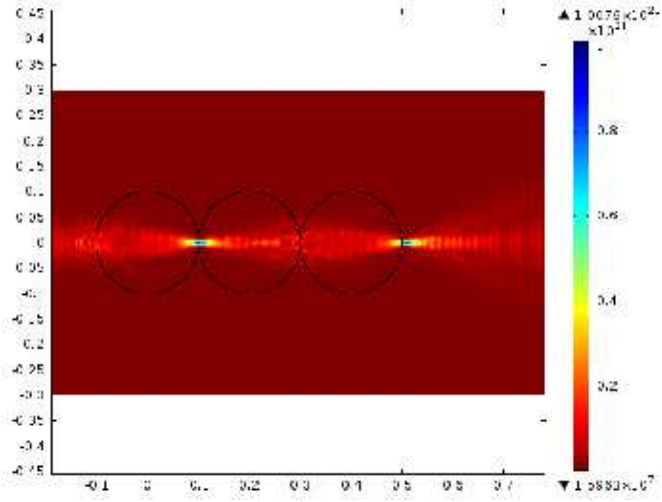}
\caption{ Linear waveguide arrangement is formed by three lenses. Simulate the propagation of the EM wave without any nonlinearity (Kerr effect). We can see that the focus points are shifted out of the surface of lenses.}
\label{fig_comsol_without_nl}
\end{figure}

As already mentioned, if the propagating wave is a Gaussian beam the focus point  shifts  outside of a single LL lens. To restore focusing one may use a nonlinear medium ~\cite{barba_3,barba_4}.
We use this approach in the LL waveguides as well.
In the Fig. \ref{fig_comsol_without_nl} we propagate a Gaussian beam through a linear Luneburg waveguide which is formed by three lenses in the absence of nonlinearity; we observe
that the focus is shifts to the exterior of the last lens.

In the presence of the  Kerr effect the refractive index is a function of the electric field  $E$, viz.
\begin{equation}
\label{kerr_index}
n(E)=n_L+\chi |E|^2
\end{equation}
where $\chi$ is the nonlinear factor with units $m^2/V^2$ and $n_L$ is the linear part of the refractive index \cite{barba_3,modern_optics,born,barba_4}.  

\begin{figure}
\centering
\includegraphics[scale=1]{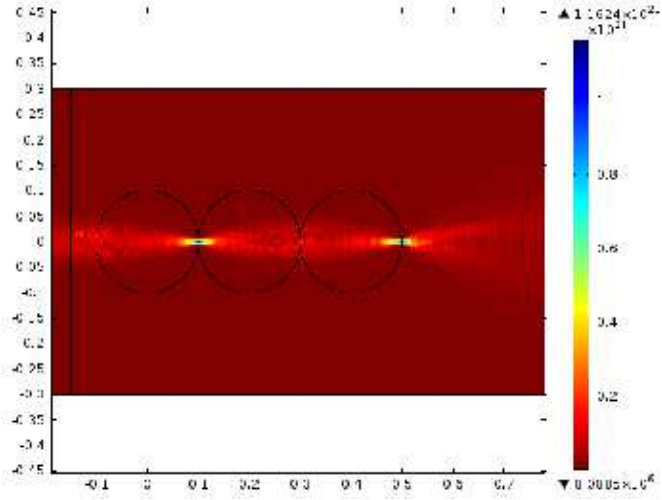}
\caption{Linear waveguide arrangement is formed by three lenses. We simulate the propagation of the EM wave including the Kerr effect. We can see that the focus points are located again on the surface of lenses.}
\label{fig_comsol_with_nl}
\end{figure}

For the numerical simulations we use a nonlinearity factor that makes the nonlinear part of the index of refraction, viz. the term $\chi |E|^2$, to be of the order of 1\% of the linear part $n_L$. In Fig. \ref{fig_comsol_without_nl} we show the shift in focusing of a LL network without nonlinear term, while in Fig. \ref{fig_comsol_with_nl} the same arrangement but with the addition of nonlinearity. We note that nonlinearity  shifts focusing in the LL network to the surface of the lenses \cite{barba_3,barba_4}.

\section{Conclusion}

Luneburg lenses  provide simple units for constructing GRIN-based metamaterials.  The advantage of using  LLs is that
one may resort to an "atomistic" picture where the lens is a single unit that has specific properties.  Subsequently,
these properties may be used in forming extended networks.   The basic feature of LLs is their property to have a 
prescribed focal surface for parallel rays that lies on the lens surface.  This property was used in the present study 
for the formation of waveguides made of Luneburg Lenses.  We saw that these guides are quite versatile since they
can channel light in an efficient way across  different paths.  

In this work we obtained an exact quasi-two dimensional solution for light trajectory through a single cylindrical LL that can be used
to study propagation in arbitrary lens arrangements.  This solution is parametrized on the initial entry point of light
on the surface of a LL as well as the initial ray angle and  gives the output location and the exit angle.  
One may thus form a type of input-output mapping that enables the study of propagation in various LL configurations.
While this solution is analytical, it fails for backward propagation; in the latter case on may resort 
multiple branches, or to a parametric two dimensional solution that is found through
a Hamiltonian approach.
  Backward propagation is now easily
handled, showing that light can follow very efficiently a full circle bend waveguide constituted of LLs.  Both types of 
solutions, viz. quasi-2D as well as true 2D give the same results in all cases compared.   The ray tracing results were compared with in house FDTD as well as commercial code (COMSOL) and
found good agreement.  For gaussian beams the LL focusing point shifts, an artifact that
can be corrected through nonlinearity also in the LL networks.  The LL networks may find usueful
applications in stronly focusing systems ~\cite{gabrielli}.

\section*{Acknowledgments} 
\noindent 
We all thank Dr. Athanasios  Gavrielides for making this collaboration
possible via the AFOSR/EORD FA8655-10-1-3039 grant.
V.K.’s work was supported via AFOSR LRIR 09RY04COR and  via the OSD
Metamaterials Insert.


\section*{Appendix}

For completion we provide in this Appendix two alternative ways to derive the LL ray tracing solution
that are based a parametric and a Helmholtz equation approach respectively. 

\appendix 
\section{Parametric 2D ray solution}

 We use the infinitesimal arc length  $ds=\sqrt{dx^2+dy^2}$ in Cartesian coordinates and further introduce the parameter $t$ as generalized time; we have $ds=\sqrt{ x'^2+ y'^2} dt$ with $ \alpha' \equiv d\alpha/ dt$ and  $x \equiv x(t)$, $y\equiv y(t)$ \cite{leonh_TO_GL, leonh_1, barba_2, barba_3, stavroudis, modern_optics,parazzoli, barba_1}. 
The Fermat integral becomes
\begin{equation}
\label{37}
\mathcal{S}=\int_A^B n(x,y)\sqrt{ x'^2+ y'^2} dt
\end{equation}
 
 where $n(x,y)$ the refractive index in cartesian coordinates; the optical Lagrangian is
\begin{equation}
\label{38}
\mathcal{L}(x,y, x', y', t)= n(x,y) \sqrt{ x'^2 + y'^2}
\end{equation}
 
We introduce  the momenta $k_x,k_y$ that are conjugate to $x,y$: 
\begin{eqnarray}
\label{m2_momentum_x}
k_x=\frac{\partial \mathcal{L}}{\partial  x'}=\frac{n  x'}{ \sqrt{ x'^2 + y'^2}} \\
\label{m2_momentum_y}
k_y=\frac{\partial \mathcal{L}}{\partial  y'}=\frac{n  y'}{ \sqrt{ x'^2 + y'^2}}
\end{eqnarray}
 
 The Eqns. (\ref{m2_momentum_x}), (\ref{m2_momentum_y}) consist of an algebraic nonlinear system with
 solution:
 \begin{equation}
 \label{m2_solution_sys}
 k_x^2+k_y^2-n(x,y)^2=0
 \end{equation}
 We  rewrite  Eq.(\ref{m2_solution_sys}) in vector using  $\vec r \equiv (x,y)$ and $\vec k \equiv (k_x,k_y)$. Subsequently
\begin{equation}
 \label{m2_solution_sys_vec}
 \vec k^2-n(\vec r)^2=0
\end{equation}

We introduce a new function $K(\vec r,\vec k)$ by
 \begin{equation}
 \label{m2_K_eq}
 K(\vec r,\vec k)=\frac{1}{2} (\vec k -n(\vec r)^2)=0
 \end{equation}
 
that has zero total differential, viz. 
\begin{equation}
dK(\vec{r},\vec k)=\frac{\partial K}{\partial \vec r}\cdot d\vec r+\frac{\partial K}{\partial \vec k}\cdot d\vec k=0
\end{equation}
We may  obtain a Hamiltonian ray tracing system by solving the Hamilton's equations for Eq.(\ref{m2_K_eq}) \cite{helmholtz,goldstein}

\begin{eqnarray}
\label{m2_dif_eq1}
\frac{d \vec r}{dt}=\frac{\partial K}{\partial \vec k}=\vec k ~~~~~~~~~~~~~~~\\
\label{m2_dif_eq2}
\frac{d \vec k}{dt}=-\frac{\partial K}{\partial \vec r}=-\frac{1}{2} \nabla n(\vec r)^2
\end{eqnarray}
where $\nabla \equiv (\frac{\partial}{\partial x},\frac{\partial}{\partial y})$ and $t$ is an effective time which is related to the real travel time  as
$dt=cd\tau	$.  Combining finally
 Eqns.(\ref{m2_dif_eq1})(\ref{m2_dif_eq2}) we obtain 
\begin{equation}
\label{m2_dif_eq_motion}
\ddot{\vec r}=\frac{c^2}{2}   \nabla n(\vec r)^2
\end{equation}
where in Eq.(\ref{m2_dif_eq_motion}) we take the derivatives respect to travel time $\tau$ thus $\dot q=dq/d\tau$ for arbitrary $q(\tau)$ \cite{leonh_TO_GL,leonh_1,stavroudis,modern_optics,born, parazzoli}.
Equation (\ref{m2_dif_eq_motion}) is a general equation  for arbitrary refractive index functions $n(\vec r)$. The explicit solution for LL will be given in the next section.

 \section{Helmholtz wave equation approach}

The stationary states of an monochromatic EM wave are given by the solutions of the 
following Helmholtz equation: 
 \begin{equation}
 \label{helmholtz}
 \left[ \nabla ^2 +(n k_0)^2 \right]  u(x,y)=0
 \end{equation}
where $u(x,y)$ is a scalar field, $n$ is the refractive index that generally depends on position,
 $k_0 \equiv \omega/c = 2 \pi /\lambda$, where $\omega$ and $ \lambda$ are the angular frequency and wavelength of the EM wave and $c$ is the velocity of light \cite{leonh_TO_GL,leonh_1,modern_optics,born,helmholtz}.

Using the well known transformation
\begin{equation}
\label{replacement}
u(x,y)=A(x,y)e^{i\phi (x,y)}
\end{equation}
 where $A,\phi$ are real, into the Helmholtz equation (\ref{helmholtz}) we obtain

\begin{eqnarray}
\label{system_eq_1}
({\nabla}\phi)^2-(n k_0)^2= \frac{\nabla^2 A}{A}\\
\label{system_eq_2}
 \nabla \cdot ( A^2 {\nabla}\phi)=0 ~~~~~~
\end{eqnarray}
 
 The last equation of the above system, Eq.(\ref{system_eq_2}), express the constancy of the flux of the vector $A^2 \nabla \phi$ along any tube formed by the field lines of the wave vector defined
 through  $\vec k ={\nabla}\phi$; the latter definition turns 
 Eq.(\ref{system_eq_1}) into
\begin{equation}
\label{eq22}
k^2-(n k_0)^2=\frac{\nabla ^2 A}{A}
\end{equation}

Using $\vec r \equiv (x,y)$ with the operator $ \nabla= \frac{\partial }{\partial \vec r} \equiv (\frac{\partial }{\partial x},\frac{\partial }{\partial y})$ and introducing
 the  function $D(\vec r,\vec k)$ 
\cite{helmholtz}.
\begin{equation}
\label{D_eq}
D(\vec r,\vec k)=\frac{c}{2k_0} \left[ k^2-(n k_0)^2-\frac{\nabla^2 A}{A} \right]
\end{equation}
we end with  a Hamiltonian ray tracing system with the following Hamilton
equations:

\begin{eqnarray}
\label{motion3_a}
 \frac{d\vec r}{dt}=\frac{\partial D}{\partial \vec k}=\frac{c\vec k}{k_0} ~~~~~~~~~~~~~~~~~~~~~~~~~~~~\\
\label{motion_3b}
\frac{d\vec k}{dt}=-\frac{\partial D}{\partial \vec r}={\nabla} \left[\frac{c k_0}{2}n^2+\frac{c}{2 k_0}\frac{\nabla ^2 A}{A} \right] 
\end{eqnarray}

The second term of the Eq.(\ref{motion_3b}) is the \textit{Helmholtz Wave Potential} \cite{helmholtz}
\begin{equation}
\label{helm_potential}
W(r) = -\frac{c}{2 k_0}\frac{\nabla ^2 A}{A}
\end{equation}

The potential of Eq.(\ref{helm_potential}) keeps the wave behavior in the ray tracing, i.e. the diffusion of the beam \cite{helmholtz}.  In the case when the space variation length $L$ of the beam amplitude $A(\vec r)$ satisfies the condition $k_0 L >> 1$ i.e. $\lambda_0 << L$, the Helmholtz potential of Eq.(\ref{helm_potential}) is approximately zero, thus the Eq.(\ref{system_eq_2}) gives the the well known \textit{eikonal equation}, which is the basic equation in geometrical optics approach \cite{luneburg,cloak,modern_optics,born,parazzoli,helmholtz}.
\begin{equation}
( \nabla \phi)^2\ = (nk_0)^2
\end{equation}
The most important result of this approach is that the rays are not coupled any more by the Helmholtz Wave Potential and they propagate independently from one another.

Finally the equations of motion take the form
\begin{eqnarray}
\label{eq_motion3_a}
\frac{d\vec r}{dt}=\frac{c}{k_0}\vec k ~~~~~\\
\label{eq_motion3_b}
\frac{d\vec k}{dt}=\frac{c k_0}{2} {\nabla}n^2
\end{eqnarray}
which are described by the Hamiltonian 
\begin{equation}
\label{m3_hamiltonian}
H(\vec r,\vec k)=\frac{c}{2k_0}\vec k^2 - \frac{c k_0}{2}n^2(\vec r)
\end{equation}

Furthermore the system of equations of motion  can be written in a second order ODE \cite{leonh_TO_GL,leonh_1,stavroudis,modern_optics,born,parazzoli}.
\begin{equation}
\label{ode3}
\ddot{\vec r}=\frac{c^2}{2} {\nabla}n^2
\end{equation}

Using now the LL refractive index of Eq.(\ref{lune_index}) we obtain the following equation of motion which gives the ray paths inside a LL.
\begin{equation}
\label{m3_dif_equation}
\ddot{\vec r}+\frac{c^2}{R^2}\vec r =0
\end{equation}

The solution of Eq.(\ref{m3_solution}) is simple; using boundary conditions $\vec{r(0)} = \vec{r_0} =(x_0,y_0)$ and $\dot{\vec{r_0}} = \vec{k_0}=(k_{0x},k_{0y})$ we obtain:
\begin{equation}
\label{m3_solution}
\left( \begin{array}{c}  x(t) \\ y(t)  \end{array} \right) =
\left( \begin{array}{c}  x_0 \\  y_0  \end{array} \right) \cos(\frac{c}{R}t)+
\left( \begin{array}{c}  k_{x0} \\  k_{y0}  \end{array} \right) \frac{R}{c} \sin(\frac{c}{R}t)
\end{equation}

The solutions (\ref{m3_solution}) in Cartesian coordinated 
describe elliptical orbits, in agreement with Luneburg's theory as well as Eq. (\ref{m1_11}).

\section*{References}



\begin{thebibliography}{99}

\bibitem{TO_meta}
A.V. Kildishev, V.M Shalaev, \textit{Transformation optics and metamaterials}, Physics-Uspekhi {\bf 54}  53-63 (2011)


\bibitem{leonh_TO_GL}
U.Leonhardt, T.Philbin, \textit{Transformation Optics and the Geometry of Light} Prog. Opt {\bf 53}, 69-152 (2009)

\bibitem{leonh_1}
Ulf Leonhardt, \textit{Notes on conformal invisibility devices}, New Journal of Physics {\bf 8} 118  (2006)

\bibitem{pendry}
D.Schurig, J.B.Pendry , D.R.Smith, \textit{Calculation of material properties and ray tracing in transformation media}, OPTICS EXPRESS {\bf 14}, 21 (2006)

\bibitem{chen}
H.Chen, C.t.Chan, P.Sheng, \textit{transformation optics and metamaterials}, Nature Materials {\bf 9} (2010) 

\bibitem{plasmons2}
Y.Liu, T.Zentgraf, G.Bartal, X.Zhang, \textit{Transformation Plasmon Optics}, Nano Letters {\bf 10}, 1991-1997 (2010)


\bibitem{luneburg}
R.K.Luneburg, \textit{Mathematical Theory of Optics}, University of California press, Berkeley and Los Angeles (1964)

\bibitem{general_solution}
S.P. Morgan, \textit{General Solution of the Luneburg Problem}, journal of applied physics {\bf 29} 9, 1358-1368 (1958)

\bibitem{falco}
A.Falco, S.C. Kehr, and U. Leonhardt, \textit{Luneburg lens in silicon photonics}, OPTICS EXPRESS {\bf 19} 6 (2011)

\bibitem{analytical_solution}
J.Sochacki, \textit{Exact analytical solution of the generalized Luneburg lens problem}, JOSA A {\bf 73} 6 (1982)

\bibitem{barba_2}
S. Takahashi, C. Chang, S. Yang, H. J. Choi, G. Barbastathis, \textit{Fabrication of Dielectric Aperiodic Nanostructured Luneburg Lens in Optical Frequencies}, in Quantum Electronics and Laser Science Conference, OSA Technical Digest (CD) (Optical Society of America, 2011), paper QTuM2

\bibitem{barba_3}
H.Gao, S.Takahashi, L.Tian, G.Barbastathis, \textit{Aperiodic subwavelength Luneburg lens with nonlinear Kerr effect compensation}, OPTICS EXPRESS {\bf 19}, 3 (2011)

\bibitem{cloak}
N.A. Mortensen, O. Sigmund, O. Breinbjerg, \textit{Prospects for poor-man's cloaking with low-contrast all-dielectric optical elements}, Journal of the European Optical Society-Rapid Publication {\bf 4} 09008 (2009)

\bibitem{plasmons1}
T.Zentgraf, Y.Liu, M.H.Mikkelsen, J.Valentine, X.Zhang, \textit{Plasmonic Luneburg and Eaton lenses}, Nature Nanotechnology {\bf 6} 3, 151-155 (2011)

\bibitem{lagrangian} 
V.Lakshminarayanan A.K.Ghatak K.Tyagarajan, \textit{Lagrangian Optics}, Kluwer Academic Publishers, Springer (2002)

\bibitem{stavroudis} O. N. Stavroudis,
\textit{The Mathematics of Geometrical and Physical Optics: The k-function and
its Ramifications}  Willey-VCH, Weinheim (2006)

\bibitem {modern_optics}
 R.Guenther, \textit{Modern Optics}, John Wiley \& Sons  (1990)

\bibitem {born}
 M.Born, E.Wolf  \textit{Principles of Optics}, Pergamon Press  (1975).

\bibitem{parazzoli}
C.G.Parazzoli, B.E.C.Koltenbah, R.B.Greegor, T.A.Lam, M.H.Tanielian, \textit{Eikonal equation for a general anisotropic or
chiral medium: application to a negative-graded index-of-refraction lens with an anisotropic material } JOSA B {\bf 23}, 439 (2006)

\bibitem{barba_4}
H.Gao, L.Tian, B.Zhang G.Barbastathis, \textit{Iterative nonlinear beam propagation using Hamiltonian ray tracing and Wigner distribution function}, OPTICS LETTERS {\bf 35} 24 (2010)




\bibitem {taflove}
 A.Taflove S.C.Hagness, \textit{Computational Electrodynamics: The Finite-Difference Time-Domain method}, Artech House Norwood MA 02062 (2005)
 
\bibitem{barba_1}
P.Stellman K.Tian G.Barbastathis, \textit{Design of Gradient Index (GRIN) Lens Using Photonic Non-Crystals}, in Conference on Lasers and Electro-Optics/Quantum Electronics and Laser Science Conference and Photonic Applications Systems Technologies, OSA Technical Digest (CD) (Optical Society of America, 2007), paper JThD121

\bibitem{helmholtz}
A.Orefice, R.Giovanelli, D.Ditto, \textit{Helmholtz wave trajectories in classical and quantum physics}, arXiv:1105.4973v3 (2011)

\bibitem {goldstein}
 H.Goldstein, C.Poole, J.Safko, \textit{Classical Mechanics}, Pearson Education (2002)


\bibitem{gabrielli}
L.H.Gabrielli, M.Lipson, \textit{Integrated Luneburg lens via ultra-strong index gradient on silicon}, OPTICS EXPRESS {\bf 19}, 21 (2011)




\end{thebibliography}
\end{document}